\documentclass[onecollarge,natbib]{svjour2}
\bibpunct{[}{]}{;}{n}{}{,} 
\smartqed 
\usepackage{graphicx}

\newcommand{\beq}{\begin{eqnarray}}
\newcommand{\eeq}{\end{eqnarray}}
\newcommand{\nneeq}{\nonumber \end{eqnarray}}

\newcommand{\nn}{\nonumber \\}
\newcommand{\es}{& = &}
\newcommand{\rs}{\, = \,}
\newcommand{\ps}{& + &}

\newcommand{\np}{\nn \ps}

\newcommand{\cM}{ {\cal M} }

\newcommand{\cU}{ {\cal U} }
\newcommand{\cL}{ {\cal L} }

\journalname{Few Body Systems}
\begin{document}
\title{ Proton radius puzzle in Hamiltonian dynamics }
\author{ Stanis{\l}aw D. G{\l}azek }
\institute{ Institute of Theoretical Physics, 
            Faculty of Physics, 
            University of Warsaw \at
            Pasteura 5, 02-093 Warsaw, Poland \\
              Tel.: +48-22-553 2824\\
              Fax:  +48-22-553-2995\\
              \email{stglazek@fuw.edu.pl}}
\date{Received: 9 October 2014 / Accepted: 21 October 2014 }

\maketitle

\begin{abstract}

Relativistic lepton-proton bound-state eigenvalue equations for
Hamiltonians derived from quantum field theory using second-order 
renormalization group procedure for effective particles, 
are reducible to two-body Schr\"odinger eigenvalue equations with 
the effective Coulomb potential that exhibits a tiny sensitivity 
to the characteristic momentum-scale of the bound system. The 
scale dependence is shown to be relevant to the theoretical 
interpretation of precisely measured lepton-proton bound-state 
energy levels in terms of a 4\% difference between the proton 
radii in muon-proton and electron-proton bound states.

\end{abstract}

\section{ Introduction }
\label{intro}

The size of proton charge distribution plays a relatively 
minor role in atomic physics since it is about five orders 
of magnitude smaller than the size of atoms. However, from
precise measurements and calculations of atomic energy levels 
a puzzling situation emerges. The proton size in the muon-proton 
bound states appears to be about 4\% smaller than in the 
electron-proton bound states. Quoting Ref.~\cite{Pohl R(2013)}, 
{\it ``...} [the radius] {\it should be a simple quantity 
to determine and understand, but that is not the case. Recent 
experimental results are not yet well understood, but future 
research may reveal the true value of this radius, lead to a 
better understanding of its structure, or demonstrate an 
unexpected aspect of its interactions.''} It is pointed out 
in this article that the front form (FF) of Hamiltonian 
dynamics~\cite{Dirac P(1949)} equipped with the renormalization 
group procedure for effective particles (RGPEP) may shed a new 
light on the issue~\cite{Glazek S(2014)}. 

Briefly speaking, the proton-size puzzle is seen here as 
emerging from an apparent discrepancy between, on the one 
hand, the Standard Model (SM) assumption that the atomic systems 
in question are ultimately described by a local quantum field 
theory (QFT) and, on the other hand, the standard atomic physics 
assumption that in the first approximation the electron-proton 
and muon-proton bound states can be described using the well-known 
two-body Schr\"odinger equation with a Coulomb potential. To be
specific, the QFT predicts that the physical systems are superpositions 
of innumerably many component states with varying and ultimately 
unlimited numbers of virtual field quanta, while the Schr\"odinger 
equation describes just two particles that interact through 
a simple potential. In essence, calculations carried out within 
the framework of QFT cannot be based on the simple Schr\"odinger 
picture without explanation of exactly how the latter is supposed 
to emerge from the former. Even if one assumes that quarks and 
gluons of QCD, as a part of the SM, form a proton that can be 
treated as nearly point-like, one is still left with a complex 
QED picture in which the lepton-proton bound states contain 
indefinite numbers of virtual photons and fermion-anti-fermion 
pairs on top of a very complex ground state, or vacuum. So far, 
the QFT vacuum complexity prevents physicists from providing 
for it any accurate construction~\cite{Dirac P(1965),Wilson et al.(1994)}.
Despite great progress in methodology and computational 
technology over more than half a century, which can be illustrated 
by many examples of highly advanced work of which only a small 
number can be quoted 
here~\cite{BetheSalpeter,GellMannLow,BetheSalpeterBook,CaswellLepage,
KinoshitaQED,Kinoshita1999,PachuckiMuon,PachuckiAtoms,Jones2,
Kinoshita2012,Mohr P(2012)}, the basic question of how a complex 
bound-state dynamics in QFT could be systematically reduced to 
the Schr\"odinger equation for just two particles still awaits
a precise answer.

The contrast between complexity of QFT and simplicity of
a few-body Schr\"odinger equation is not specific to atomic
physics. Even more striking contrast exists between the complexity 
of QCD and simplicity of the constituent quark model, a basis for 
classification of hadrons which says that mesons are made of 
two and baryons of three constituents. However, a full-fledged 
application of the RGPEP to a complete version of the Standard 
Model, or its potential extensions, is a long way off. 

This article is very limited in scope. It only concerns 
corrections to lepton-proton ground-state energy levels 
due to the proton radius. The point is that these corrections 
are sensitive to the effective nature of the two-body 
Schr\"odinger equation and the sensitivity appears large 
enough for taking it into account in extracting the proton 
radius from the measured energy levels. However, it should 
be stressed that many other corrections due to the effective 
nature of the two-body Schr\"odinger equation need to be 
calculated first before one can precisely establish the 
magnitude of small energy terms that depend on the proton 
radius.

In order to outline how the FF of Hamiltonian dynamics 
equipped with the RGPEP may help in resolving the proton 
radius puzzle, the article is organized in the following 
way. Sec.~\ref{protonsize} explains what is meant by the 
proton radius puzzle. The RGPEP derivation of the 
Schr\"odinger equation from QFT is outlined in Sec.~\ref{path}
and key details of the derivation are described in Sec.~\ref{derivation}.
Sec.~\ref{radius} indicates the possibility of a resolution 
of the proton radius puzzle using the RGPEP. Sec.~\ref{C} 
concludes the article with general comments concerning the 
proton radius puzzle, Schr\"odinger equation for few-body 
systems and QFT.

\section{ Puzzle of the proton size in the Schr\"odinger equation }
\label{protonsize}

The first-approximation Schr\"odinger eigenvalue equation 
for a lepton-proton system internal dynamics in momentum 
representation has the form (using the convention that
$\hbar = c = 1$)
\beq
\label{Hpsi=Epsi}
{ \vec p^{\,2} \over 2\mu} \, \psi(\vec p )
+ \int {d^3 k \over (2\pi)^3 } \, 
V(\vec p, \vec k ) \, \psi(\vec k )
\es 
-E \, \psi(\vec p ) \ ,
\eeq
where $\vec k$ and $\vec p$ denote the relative momentum
of the lepton with respect to the proton, $\mu$ is the 
reduced mass and $E$ is the binding energy. The 
interaction kernel $V(\vec p, \vec k )$ is the 
spin-independent Coulomb potential for point-like 
charges,
\beq
V(\vec p, \vec k ) 
\es
V^{\rm pt}_C(\vec q ) 
\rs
- \ { 4 \pi \alpha \over \vec q^2 } \ ,
\eeq
where $\vec q = \vec p - \vec k$ is called the momentum 
transfer and $\alpha \sim 1/137$ is the fine structure 
constant. The extended proton charge distribution is 
accounted for by replacing $V^{\rm pt}_C(\vec q)$ with
\beq
\label{VG}
V_C(\vec q \,) 
\es
V^{\rm pt}_C(\vec q ) \ G_E(\vec q^2) \ ,
\eeq
where $G_E(\vec q^2)$ is the proton electric form factor. 
For the small values of momentum transfer that characterize 
atomic systems,
\beq
\label{GE}
G_E(\vec q^2) = 1 - {1\over 6} \, r_p^2 \, \vec
q^2 + o(\vec q^2) \ ,
\eeq
where $r_p$ denotes the proton radius. This radius is
understood to be a physical property of the proton 
and it is expected to be the same in the muon-proton
and electron-proton bound states. Keeping only the first 
two terms in the expansion, one obtains the corrected Coulomb 
potential in the form
\beq
\label{correction1}
V_C(\vec q ) 
\es
V^{\rm pt}_C(\vec q ) + \delta V(\vec p, \vec k ) \ ,
\eeq
where the correction term is
\beq
\label{VCrpp}
\delta V(\vec p, \vec k ) \es  { 2\pi \alpha \over 3} \, r_p^2 \ .
\eeq
The corresponding correction to the Coulomb potential 
in position representation, i.e., a correction to 
$-\alpha/| \vec r |$, is           
\beq
\label{VCrpr}
\delta V( \vec r )
\es
{ 2\pi \alpha \over 3} \, r_p^2 \ \delta^3(\vec r ) \ .
\eeq
Thus, the first-order corrections to energy levels due
to the extended nature of the proton charge distribution 
are described by the formula
\beq
\label{IntroDeltaEda}
\Delta E \es { 2\pi \alpha \over 3} \ r_p^2 \ |\hat \psi(0)|^2  \ ,
\eeq
where the wave function at $\vec r = 0$ is
\beq
\hat \psi(0) \es \int {d^3k \over (2\pi)^3 } \ \psi(\vec k ) \ .
\eeq
For example, in the ground state described by 
Eq.~(\ref{Hpsi=Epsi}), where
$
|\hat \psi(0)|^2  
=
(\alpha \mu)^3  / \pi $,
the correction due to the proton radius is
\beq
\label{DeltaEda11}
\Delta E \es 
{ 4 \over 3} \ 
(\alpha \mu \, r_p)^2 \ 
{ \mu \alpha^2  \over 2} \ .
\eeq
The factor $\alpha \mu$ in the electron-proton bound
state is practically equal to the inverse of the Bohr 
radius, $r_B$. So, the product $\alpha \mu \, r_p$ is the 
ratio of the proton and hydrogen radii, $r_p/r_B \sim 10^{-5}$. 
Hence, the energy correction $\Delta E$ is on the order
of $10^{-10}$ times Rydberg. In the muon-proton bound state, 
the reduced mass is about 200 times greater than in the 
electron-proton bound state, because muons are about 200 
times heavier than electrons. Therefore, the correction 
to energy is $200^3$ times larger and precise energy 
measurements allow one to extract the value of the proton 
radius, $r_p$, from muon-proton bound-state data with accuracy 
reaching 1\%~\cite{Bernauer et al.(2010),Pohl R et al.(2010)}. 

Of course, many other corrections need to be included in 
the calculations for atomic systems before one can separate 
the corrections due to the small proton radius. Such 
calculations have a long history of extensive work using
different methods~\cite{BetheSalpeter,GellMannLow,BetheSalpeterBook,
CaswellLepage,KinoshitaQED,Kinoshita1999,PachuckiMuon,PachuckiAtoms,
Jones2,Kinoshita2012,Mohr P(2012)}. The proton radius puzzle is 
that, after all the calculated terms are included, the quantity 
$r_p$ extracted from the muon-proton bound-state data is smaller 
by about 4\% than the corresponding quantity extracted from the 
electron-proton bound-state and scattering data (with an exception 
of some dispersion analysis of electron-proton scattering data, 
such as in  Ref.~\cite{Belushkin M(2007)}, discussed in 
Ref.~\cite{Pohl R(2013)}). This article deals exclusively with 
the energy correction $\Delta E$ and focuses on the changes 
in its interpretation that follow from the effective nature of 
the two-body Schr\"odinger equation for lepton-proton bound states.

\section{ Outline of the RGPEP path from QFT to the Schr\"odinger equation }
\label{path}

The RGPEP allows one to derive the Schr\"odinger equation
for two-body bound states from relativistic QFT in several 
steps~\cite{Glazek S(2014)}. One begins by proposing a 
classical field-theoretic Lagrangian density that formally 
leads to an associated canonical FF Hamiltonian density, 
accounting for constraints. Then one quantizes the independent
field degrees of freedom and integrates the resulting quantum 
density over the front to obtain the Hamiltonian operator.  
This operator is singular. One regulates it to remove the
infinities in favor of the presence of cutoffs. The 
regulated canonical Hamiltonian provides a trial initial
condition for solving the first-order differential equation
of the RGPEP. 

The differentiation is defined with respect to the RGPEP 
momentum scale parameter $\lambda$, which labels members 
of a family of effective Hamiltonians. The scale parameter 
$\lambda$ varies from infinity, which labels the regulated 
canonical Hamiltonian, down to the characteristic momentum 
scale of the phenomena one is interested in describing,
which labels the effective Hamiltonian that provides the 
simplest dynamical explanation of the phenomena of interest. 
The observables do not depend on $\lambda$, by construction, 
but the effective degrees of freedom and their dynamics do.

The generic QFT difficulty is that the initial condition 
provided by the regulated canonical Hamiltonian yields 
solutions for the effective Hamiltonians that depend on 
the regularization cutoff parameters in diverging ways. 
Therefore, the initial condition must be modified until 
the required effective Hamiltonian ceases to depend on 
the regularization. The modification is achieved by 
finding suitable counter-terms that are added to the 
initial condition. The conceptual difficulty is that 
the conditions that determine the counter-terms concern
the effective Hamiltonian with a finite value of $\lambda$
while the counter-terms are only inserted in the initial-condition 
Hamiltonian at $\lambda \to \infty$. However, when the coupling 
constant is as small as $\alpha \sim 1/137$ and the issues of 
confinement are not overwhelmingly important (proton can be 
treated as nearly elementary particle), one can solve the 
differential RGPEP equation using an expansion in powers of 
$\alpha$ and find the counter-terms order by order.

The effective Hamiltonian with a suitable $\lambda$ for 
lepton-proton bound-state calculations is relatively easy 
to calculate in the relevant QFT with accuracy up to terms 
of order $\alpha$~\cite{Glazek S(2014)}. Such accuracy is 
sufficient to derive the effective interaction that corresponds 
to the Coulomb potential in the two-body Schr\"odinger 
equation, Eq.~(\ref{Hpsi=Epsi}), with an extended proton charge.
The non-local nature of the proton charge distribution can 
be accounted for by inserting the proton electric form factor 
in the initial-condition Hamiltonian at $\lambda = \infty$ 
and including it in all steps of the entire procedure. The 
eigenvalue problem for the resulting Hamiltonian still involves 
infinitely many Fock components because the terms of formal 
order $\sqrt{\alpha}$ are capable of producing photons from 
fermions and fermion-anti-fermion pairs from photons. But 
these components are now built using the effective-particle 
creation operators characterized by a finite scale $\lambda$, 
instead of the ones for bare quanta characterized by $\lambda 
= \infty$. Moreover, the effective-particle creation operators 
are now applied to a simple vacuum state. 

The elimination of vacuum difficulty is a characteristic 
feature of the regulated FF Hamiltonian dynamics, e.g., 
see Ref.~\cite{Wilson et al.(1994)}. The change from bare 
to effective particles, is a new feature brought in by 
the RGPEP. The interaction vertices in the Hamiltonian for 
effective particles contain exponential form factors of 
width $\lambda$ in momentum variables. These form factors
are obtained by solving the RGPEP equations. They limit 
the allowed changes of the invariant mass of interacting 
effective particles. Therefore, the couplings 
between sectors with different numbers of particles are 
suppressed at large energies and cannot produce large 
effects if the coupling constant is small. In case of a
lepton-proton bound state, one can assume that its dominant
component is the state of just one effective lepton and one
effective proton. The dynamical contribution of the effective 
lepton-proton-photon component can be calculated to order 
$\alpha$. In summary, the great complexity of QFT is absorbed 
in the internal structure of the two effective constituents
that interact through an effective potential.

The resulting effective two-body eigenvalue problem is nearly 
identical to Eq.~(\ref{Hpsi=Epsi}), except that the Coulomb 
interaction is modified by the RGPEP form factor of width 
$\lambda \sim \sqrt{ \mu m}$, where $\mu$ is the reduced and 
$m$ the average mass of constituents. It turns out that this 
form factor is capable of generating a few-percent effect in 
the tiny corrections to atomic energy levels from which the
the proton radius is extracted. 

\section{ Key details of a derivation of the Schr\"odinger equation from QFT }
\label{derivation}

The Lagrangian density we start from is 
\beq
\cL \es - {1 \over 4} F_{\mu \nu} F^{\mu \nu}
        + \sum_{n=1}^3 \bar \psi_n( i \partial \hspace{-5pt}/ 
        - e_n A \hspace{-5pt}/ - m_n) \psi \ ,
\eeq   
where the subscripts $n=1, 2, 3$ refer to electrons, 
muons and protons, correspondingly, and at this stage 
protons are still considered point-like. The proton 
form factor is inserted at a later stage, in the FF 
Hamiltonian that we derive starting from $\cL$. All 
tensors are written using the same convention that is 
used for Minkowski's space-time coordinates, 
$x^\pm  = x^0 \pm x^3$ and 
$x^\perp = (x^1, x^2)$.
Variable $x^+$ plays the role of time and $x^\perp$ and 
$x^-$ play the roles of space co-ordinates. In gauge 
$A^+=0$, the canonical FF Hamiltonian is~\cite{Yan T(1973)}
\beq
\label{Pminus}
P^- 
\es  
\label{T+-4abbreviated}
\int dx^- d^2x^\perp \,
\left\{
{1 \over 2} A_\mu  \, \partial^{\perp \, 2} A^\mu
+ 
\sum_{n=1}^3
\left[
\bar \psi_n \gamma^+  
{ - \partial^{\perp \, 2} + m_n^2 \over 2 i \partial^+} \, 
\psi_n
+
e_n \, \bar \psi_n A \hspace{-5pt}/ \, \psi_n
\right.
\right.
\np
\left.
\left.
e_n^2 \, 
\bar \psi_n A \hspace{-5pt}/ \, 
         {\gamma^+ \over 2i \partial^+} \,
          A \hspace{-5pt}/ \, \psi_n
+
e_n \bar \psi_n \gamma^+ \psi_n \,          
{ 1 \over 2( i\partial^+)^2 } \,
\sum_{k=1}^3 \, e_k\,
\bar \psi_k \gamma^+ \psi_k 
\right]
\right\}
\, , \nn
\eeq  
where the dependent components of fields,
$A^-$ and $\psi_{n-}$, are solutions to the
constraint equations with electric charge 
set to zero. Quantization is carried out by replacing 
the independent fields $A^\perp$ and $\psi_{n+}$ 
by operators $\hat A^\perp$ and $\hat \psi_{n+}$
expanded into their Fourier modes that satisfy
the standard FF commutation relations. The 
quantum Hamiltonian is regulated by limiting
the range of momenta in the Fourier expansions
of the fields or by limiting the changes of
momenta in the interaction terms~\cite{Glazek S(1999)}.
The required counter-terms up to order $\alpha$ 
are found in the process of evaluating the Hamiltonian 
with a finite momentum scale $\lambda$, using the 
condition that its matrix elements in the basis 
states of small invariant mass do not depend 
on the regularization. The scale $\lambda$ is 
introduced by the unitary rotation of particle 
creation and annihilation operators, commonly 
denoted below by $q$, from the bare ones at $\lambda = \infty$
to the effective ones at a finite $\lambda$,
\beq
\label{qlambda}
q_\lambda \es \cU_\lambda \, q_\infty \, \cU_\lambda^\dagger \, ,
\eeq
and keeping the Hamiltonian operator unchanged,
\beq
\label{cHt}
H_\lambda(q_\lambda) \es H_\infty(q_\infty) \, .
\eeq
Differentiation with respect to $\lambda$
yields the RGPEP evolution equation, which includes
a Hamiltonian-dependent kernel designed to decrease 
the range of allowed off-shellness in interaction 
terms as $\lambda$ decreases. This equation is solved
using expansion in powers of $\alpha$. Here it is 
sufficient to include the terms of formal order 1, 
$\sqrt{\alpha}$ and $\alpha$. The calculation follows 
the same steps as in the case of heavy 
quarkonia~\cite{Glazek S(2004)}, but it is much 
simpler because the theory is simpler than QCD. For 
example, the lepton and proton mass-squared counter-terms 
can be adjusted to physical fermion mass values using 
the single fermion eigenvalue equations for $H_\lambda$, 
since the fermions are not confined (see \cite{Glazek S(2014)}).

Having established the effective Hamiltonian of
scale $\lambda$ up to order $\alpha$, one poses its
eigenvalue problem for lepton-proton bound states. 
The eigenstates are written in terms of a basis in 
the Fock space built using the creation operators 
corresponding to $\lambda$. The expansion into
the effective Fock components is exponentially 
suppressed for invariant-mass changes greater than
$\lambda$. Therefore, for very small $\alpha$,
one can approximate the bound state by a superposition 
of the effective lepton-proton and lepton-proton-photon 
components. This is sufficient to derive the equivalent 
effective eigenvalue problem of accuracy up to terms of 
formal order $\alpha$ in the effective two-body lepton-proton 
component, accounting for the three-body component via a 
one effective-photon exchange between the effective 
fermions (the associated self-interaction terms are 
accounted for in the mass-squared terms). The proton 
electric form factor inserted in the initial Hamiltonian
appears unchanged in the resulting eigenvalue equation 
\beq
\label{Hpsi=EpsiRGPEP}
{ \vec p^{\,2} \over 2\mu} \, \psi(\vec p\,)
+ \int {d^3 k \over (2\pi)^3 } \, 
V_\lambda(\vec p, \vec k\,) \, \psi(\vec k\,)
\es 
-E_B \, \psi(\vec p\,) \ ,
\eeq
where
\beq
\label{f}
V_\lambda(\vec p, \vec k\,)
\es 
f_\lambda(\vec p, \vec k )
\
V_C^{\rm pt}(\vec q\,) \ G_E(\vec q^2) \ .
\eeq
The interaction differs from $V(\vec p, \vec k\,)$ 
in Eq.~(\ref{VG}) by the scale dependent RGPEP 
form factor 
\beq
f_\lambda(\vec p, \vec k ) \es e^{- (\Delta \cM^2/\lambda^2)^2} \ ,
\eeq
in which $\Delta \cM^2$ is the difference of lepton-proton 
free invariant masses squared before and after the interaction,
\beq
\Delta \cM^2 
\es
\cM^2(\vec p) - \cM^2(\vec k) \ .
\eeq
For a lepton of mass $m_l$ and proton of mass $m_p$, 
one has~\cite{Glazek S(2014)}
\beq
\cM^2(\vec p) \es
(m_l + m_p)^2\left[ 1 + \vec p^2/(m_l m_p) \right] 
\eeq
and the RGPEP form factor reads
\beq
f_\lambda(\vec p, \vec k ) 
\es 
e^{ - ( \vec p^2 - \vec k^2)^2 (m_l + m_p)^4/(\lambda^2 m_l m_p)^2} \ .
\eeq
To match the well-known universal scaling of the 
atomic Schr\"odinger equation with $\alpha$, one 
needs
\beq
\label{lambda}
\lambda = a \, \sqrt{ \mu \, (m_l + m_p)/2 }  \ ,
\eeq
where the number $a$ is not expected to differ 
a lot from 1. For the Schr\"odinger equation 
is known to be valid in the whole range of momenta 
smaller than the constituent masses irrespective 
of the values of the masses and it does not involve 
important contributions from the range of momenta 
much larger than the masses. Thus, the form factor 
$f_\lambda(\vec p, \vec k )$ that changes the abstract 
Coulomb potential for point-like particles, 
Eq.~(\ref{VG}), to the effective potential that 
befits the effective nature of constituents in 
the Schr\"odinger equation, Eq.~(\ref{f}), has the 
form 
\beq
\label{fBohr}
f_\lambda(\vec p, \vec k ) 
\es 
e^{ - ( \vec p^2 - \vec k^2)^2 /(a \mu)^4} \ .
\eeq

\section{ Toward a solution of the proton radius puzzle using the RGPEP }
\label{radius}

Using the input from Secs.~\ref{protonsize}, 
\ref{path} and \ref{derivation}, one arrives 
at the conclusion that the small two-body 
interaction Hamiltonian term due to the physical 
proton radius is not given by Eq.~(\ref{VCrpp}), 
but by the formula
\beq
\label{radiuscorrection}
\delta V(\vec p, \vec k ) \es  f_\lambda(\vec p, \vec k ) \ 
{ 2\pi \alpha \over 3} \, r_p^2 \ .
\eeq
Consequently, the associated energy corrections 
are described not by Eq.~(\ref{IntroDeltaEda}), 
but by the formula 
\beq
\label{DeltaEda}
\Delta E \es c_\lambda \ { 2\pi \alpha \over 3} \ r_p^2 \ |\hat \psi(0)|^2  
\eeq
where the coefficient $c_\lambda$ is given by
\beq
c_\lambda
\es
{ 1 \over |\hat \psi(0)|^2 } 
\int {d^3p \over (2\pi)^3}
\int {d^3k \over (2\pi)^3}
\  \psi(\vec p) \
f_\lambda(\vec p , \vec k )
\ \psi(\vec k) \ ,
\eeq
for the wave functions normalized to 1.
Since $f_\lambda < 1$ for most values of
the momenta, the result for $c_\lambda$ 
is smaller than 1.

The magnitude of deviation of $c_\lambda$ from 1 
can be estimated using the momentum variables 
in units of the Bohr momentum, or $\vec p = 
\alpha \mu \vec u$ and $\vec k = \alpha \mu \vec 
v$. In these variables, Eq.~(\ref{fBohr}) reads 
\beq
\label{fBohr1}
f_\lambda(\vec p, \vec k ) 
\es 
e^{ - (\alpha/a)^4 \, ( \vec u^2 - \vec v^2)^2} \ .
\eeq
The exponential contains $(\alpha/a)^4$ and one 
might think that $c_\lambda$ differs from 1 by 
terms order $(\alpha/a)^4 \sim 3 \cdot 10^{-9}/a^4$,
which would be a negligible effect. Thus, if one
assumed that an expansion in powers of $\alpha$
should apply, the effective nature of the 
Schr\"odinger equation could be ignored. However, 
the expansion in powers of $\alpha$ does not apply. 
For example, the unperturbed ground-state wave functions  
are $N/(1+ \vec u^2)^2$ and $N/(1+ \vec v^2)^2$,
where $N$ is the normalization constant. Therefore, 
the term order $\alpha^4$ in expansion of $c_\lambda$ 
is divergent. One has to evaluate $c_\lambda$ numerically. 
Since muons are about 200 times heavier than electrons, 
the same value of $\lambda$ in one and the same effective
theory for both electron-proton and muon-proton systems 
implies that the factor $a$ in Eq.~(\ref{lambda}) is 
about 14 times smaller in the muon-proton system than 
in the electron-proton system. Such large change of $a$ 
is capable of changing $c_\lambda$ from nearly 1 in the 
electron-proton system to about 0.92 in the muon-proton 
system. The associated 8\% reduction in the radius squared 
means 4\% reduction in the radius itself, which is 
about the magnitude of the difference that causes
the proton radius puzzle. 

\section{ Conclusion } 
\label{C}

According to the RGPEP, the proton radius puzzle
stems from the difference between Eqs.~(\ref{IntroDeltaEda})
and (\ref{DeltaEda}). If one employs the usual 
Eq.~(\ref{IntroDeltaEda}), the proton radius is 
extracted from small corrections to energy levels
as the quantity which, according to Eq.~(\ref{DeltaEda}), 
is $c_\lambda r_p^2$, instead of $r_p^2$ itself. The 
coefficient $c_\lambda$ accounts for the effective
nature of constituents that appear in the two-body 
Schr\"odinger equation for lepton-proton bound states.
Being smaller than 1, $c_\lambda$ implies that the usual 
way of interpreting the standard two-body Schr\"odinger 
equation leads to extraction of a slightly smaller 
quantity for the proton radius than its value in the 
electric form factor; the heavier the lepton the smaller 
the extracted quantity. The numerical deviation of 
$c_\lambda$ from 1 is of the order required for 
solving the puzzle.

However, it would be premature to conclude that the 
RGPEP factor $c_\lambda$ has already solved the proton 
radius puzzle. The reason is that there are many other 
corrections to the energy levels known in other 
approaches that need to be calculated using the RGPEP
before it could reliably help in isolating the size of 
corrections due to the proton charge radius and in
extracting the latter from atomic and scattering data. 
Carrying out such calculations constitutes, on the one 
hand, a major research program and, on the other hand, 
presents itself as an opportunity for the theory of 
few-body systems to assert its position in the field 
of particle dynamics, not only in the atomic physics 
but in all areas of physics where equations of the 
Schr\"odinger type apply as approximate representations
of the underlying theory.

\end{document}